\documentclass[12pt]{article}
\usepackage{amsmath}
\usepackage{graphicx,soul,color}
\usepackage{enumerate}
\usepackage{natbib}
\usepackage{url} 
\usepackage{inputenc}
\usepackage[english]{babel}
\usepackage{amsthm}
\newtheoremstyle{mystyle}
  {}
  {}
  {\itshape}
  {}
  {\bfseries}
  {.}
  { }
  {}

\theoremstyle{mystyle}
\newtheorem{theorem}{Theorem}
\newtheorem*{corollary}{Corollary}
\newcommand{\blind}{1}

\begin{document}

\def\spacingset#1{\renewcommand{\baselinestretch}%
{#1}\small\normalsize} \spacingset{1}


\if1\blind
{  
  \title{\bf Evidence synthesis with reconstructed survival data}
 \author{Chenqi Fu \hspace{.2cm} \\
    Department of Public Health Sciences,\\ Pennsylvania State University \vspace{.2cm} \\ 
    Shouhao Zhou\thanks{corresponding author: Shouhao Zhou, shouhao.zhou@psu.edu} \hspace{.2cm}\\
    Department of Public Health Sciences,\\ Pennsylvania State University \vspace{.2cm} \\ 
    Xuelin Huang \hspace{.2cm}\\
    Department of Biostatistics,\\ The University of Texas MD Anderson Cancer Center\vspace{.2cm} \\
    Nicholas J. Short \hspace{.2cm}\\
    Department of Leukemia,\\ The University of Texas MD Anderson Cancer Center\vspace{.2cm} \\
    Farhad Ravandi-Kashani \hspace{.2cm}\\
    Department of Leukemia,\\ The University of Texas MD Anderson Cancer Center \\
    and \\
    Donald A. Berry \hspace{.2cm}\\
    Department of Biostatistics,\\ The University of Texas MD Anderson Cancer Center }
  \maketitle
} \fi

\if0\blind
{
  \bigskip
  \bigskip
  \bigskip
  \begin{center}
    {\LARGE\bf Evidence synthesis with reconstructed survival data}
\end{center}
  \medskip
} \fi

\bigskip\newpage
\begin{abstract}

We present a general approach to synthesizing evidence of time-to-event endpoints in meta-analyses of aggregate data (AD). Our work goes beyond most previous meta-analytic research by using reconstructed survival data as a source of information. 
A Bayesian multilevel regression model, called the ``meta-analysis of reconstructed survival data'' (MARS), is introduced, by modeling and integrating reconstructed survival information with other types of summary data, to estimate the hazard ratio function and survival probabilities. 
The method attempts to reduce selection bias, and relaxes the presumption of proportional hazards in individual clinical studies 
from the conventional 
approaches restricted to hazard ratio estimates.  
Theoretically, we establish the asymptotic consistency of MARS, and investigate its relative efficiency with respect to the individual participant data (IPD) meta-analysis.
In simulation studies, the MARS demonstrated comparable performance to IPD meta-analysis with minor deviation from the true values, suggesting great robustness and efficiency achievable in AD meta-analysis with finite sample. Finally, we applied MARS in a meta-analysis of acute myeloid leukemia to assess the association of minimal residual disease with survival, to help respond to FDA's emerging concerns on translational use of surrogate biomarker in drug development of  hematologic malignancies.
\end{abstract}

\noindent%
\vfill

\newpage
\spacingset{1.9} 
\section{Introduction}
\label{sec:intro}

The time to occurrence of a disease-related event is the most important outcome measure in clinical trials to assess the efficacy of an intervention.
Because the clinical and regulatory evaluation of a new intervention usually involves multiple trials, evidence synthesis using meta-analysis that integrates the survival data of such trials is essential \citep{FDA2018}. 

In systematic reviews, aggregate data (AD) meta-analysis is the mainstay approach to combining the results of multiple trials. Even though individual patient data (IPD) meta-analysis is considered the gold standard \citep{Stewart1995, Riley2010}, IPD were hard to obtain \citep{Tierney2015, Marciniak2017}.  A recent systematic survey showed that less than $1\%$ of 78,629 meta-analyses included complete or partial IPD, while the vast majority were based on AD only \citep{Nevitt2017}.  Thus, it is of utmost importance to develop sensible methodology for AD meta-analysis. 

We are concerned in this work with AD meta-analytic methods to synthesize evidence in time-to-event data. For survival outcomes, the standard approach entails a fixed or random effects model
on aggregated study-level hazard ratio (HR) estimates \citep{Moher2010, Higgins2020}.  
Owing to data censoring in studies of survival outcomes, HRs are useful for describing differences in treatment effects between an intervention group and control group.
Abundant open-source and commercial computational software programs are also conveniently available for conducting meta-analysis.   

However, this standard approach poses several methodological limitations and challenges in analysis and reporting. 
First, applications are constricted under the assumption of proportional hazards.  Recent systematic reviews of randomized controlled trials showed that the proportional hazards assumption could be significantly violated in one-fifth of treatment comparisons \citep{Rulli2018}, while testing of the proportional hazards assumption seldom was conducted \citep{batson2016review}. In fact, significant violations, such as survival curve convergences and crossings, are common in clinical studies,  which gives rise to concerns upon general model validity
of AD meta-analysis for time-to-event outcomes  \citep{Kristiansen2012}. 

Second, the exclusive use of HR estimates impairs the effectiveness and generalizability of meta-analysis. The strength of systematic review is that it allows evidence to be synthesized from all available clinical information in the literature and allows meta-analysis to combine the information comprehensively.
Using HR estimates as the only source of ``survival data'' 
simplifies analyses but undermines the generalizability of results by excluding studies that alternatively report other kinds of survival data, such as Kaplan-Meier (KM) curves, log-rank test statistics, median survival times or survival rates.
Existing methods to convert those types of survival data to HR estimates involve additional assumptions about time-to-event outcomes \citep{Tudur2001, Tierney2007} and, given the various duration of studies, could assign study weights improperly  \citep{riley2007evidence}.

The last limitation is pertinent to AD meta-analytic reporting for intervention effect. 
If the proportional hazards assumption is violated in a single study, the interpretation of the overall meta-analytic HR statistics could be a problem.
Moreover, it is preeminent to determine survival 
curves, rather than a hazard ratio, 
to gauge the risk of the event over time \citep{pocock2002survival}. 
Although various methods have been developed in the literature for combining published survival curves \citep{begg1989bone,voest1989meta,shore1990meta,reimold1992assessment,hunink1994meta,dear1994iterative,Arends2008}, they were all rooted in the proportional hazards assumption except \cite{combescure2014meta}. That approach applies a distribution-free approach to estimating the summary survival curve for single-arm studies, but  is subject to a risk of bias when multiple curves from the same study are modeled. In brief, AD meta-analytic reporting that entails comparative trials without the proportional hazards assumption remains unclear.

The main contribution of this paper is to address these challenges and present a novel AD meta-analysis framework for survival-type endpoints. 
Particularly, a Bayesian multilevel modeling strategy, coined ``meta-analysis of reconstructed survival data'' (MARS), is proposed to exploit reconstructed survival curves as a richer source of survival information.
In general, survival data can be reconstructed by 1) acquiring the positions of censoring and event points on images of KM survival curves; 2) performing iterative numerical computation to convert 2-dimensional position data into time-to-censoring and time-to-events.  The use of the reconstructed survival data, which can be regarded an alternative to IPD, enables several advantages similar to using IPD in both modeling and reporting.  For example, instead of assuming proportional hazards, MARS can estimate the time-varying hazard rates and HRs. 
We note that KM curves may not be provided in every published clinical study.  To conduct a meta-analysis that thoroughly covers the literature, MARS also simultaneously accommodates 2 more types of AD: HR estimates and survival rates at various time points.


Our goal is to deliver effective estimation for the comparison of efficacy of treatments, and thus expand the traditional framework for AD meta-analysis.
The results of simulation study suggest that MARS is competent to provide valid estimation of a treatment effect in both cases of constant HR and time-varying HR, and capable of reporting informative summary metrics, such as survival probabilities, median survival times and restricted mean survival times (RMSTs). Moreover, MARS generates HR estimates that are consistent and have decent asymptotic efficiency with respect to IPD meta-analysis.

The outline of the paper is as follows.  In Section 2, we present a data set in adult leukemia with disease-free survival (DFS) as the time-to-event outcome of interest.  Our work was originally motivated by a larger AD study consisting of pediatric and adult leukemia patients with DFS endpoints. Here we apply a subset of 64 non-transplantation trials to illustrate the concept. In section 3, we introduce the MARS, which can integrate various types of survival information, and discuss the metrics to quantify treatment effects for informative reporting. The theoretical properties of MARS are investigated in section 4. Simulation studies are illustrated in section 5 for finite sample performance and an application of MARS to leukemia data set is demonstrated in section 6. Lastly, section 7 provides a discussion of our work, with suggestions for further research.

\section{Case study: minimal residual disease in leukemia}
\label{sec:2}

Responding to the emerging development of targeted and immunological treatments, the U.S. Food and Drug Administration \citep{us2019hematologic} published a guideline clarifying for the use of minimal residual disease (MRD) status in the development of drugs for the treatment of hematologic malignancies. An MRD-positive status means that the disease was still detectable after treatment, whereas an MRD-negative status means that no disease was detected after treatment. Potentially, the MRD status can serve as a prognostic biomarker to assess a patient’s risk of future relapse for regulatory and clinical uses. 

Evidence supporting the clinical validity of MRD status as a surrogate biomarker varies across hematologic cancer types and patient populations \citep{rach2016regulatory, Berry2017}. To gain a better understanding of the state-of-the-science of the use of MRD for specific hematologic malignancies, the FDA recommended a systematic review and meta-analysis. In particular, to ensure the effective validation of MRD to support the marketing approval of drugs and biological products, the FDA gave special meta-analytic considerations to (i) prespecified inclusion criteria to ensure comprehensive literature search of both positive and negative results, to limit selection bias, and (ii) sensitivity analyses (e.g., alternative statistical methods) to demonstrate the robustness of MRD as a surrogate marker.  

\citet{short2020association} performed a quantitative overview of clinical studies in pediatric and adult patients with acute myeloid leukemia (AML) to quantify the relationships between disease-free survival (DFS) and MRD status.
If a strong association can be demonstrated with robustness across studies, MRD status could be used as a critical indicator of therapeutic benefit in clinical practice to guide the clinical treatment of AML patients and accelerate the drug development by providing early evidence of treatment benefit in regulatory approval.
After performing a comprehensive literature search and screening based on inclusion and exclusion criteria, we took sequential steps to extract various survival information from articles selected in the meta-analysis.  
First, survival data were reconstructed from the KM estimators provided in 16 studies. Second, we accepted the hazard ratios reported in the articles comparing MRD positive group to MRD negative group if the survival data cannot be reconstructed. This consists of 36 studies reported log HRs and corresponding standard errors.  
Finally, 12 additional studies provided only survival rates at particular time points, which were still included as a type of supplemental survival data in the meta-analysis. In total, we included 64 studies with 8,033 AML patients who did not undergo transplantation to illustrate MARS (proposed in section 3) and to respond to the FDA’s pertinent concerns.

\section{Model for survival AD meta-analysis}
\label{sec:3}

We present a general framework for survival meta-analysis, called the “meta-analysis of reconstructed survival data” (MARS). MARS attempts to establish a one-stage Bayesian multilevel model with separate AD components, such that the most informative summary survival data available in individual studies can be properly integrated to the parameter estimation and uncertainty quantification. 

In this section, we first review the general strategy to reconstruct the survival data from KM curves, and discuss the model specification 
to relax the proportional hazards assumption.  If KM curves in a decent quality are unavailable in some clinical studies, we elicit and incorporate other types of survival data, though less informative, to complement a comprehensive AD meta-analysis and diminish selection bias.  We provide the hierarchical prior distributions to complete the model specification, and finally present several alternative summary measures from MARS to better inform AD meta-analysis reporting.


\subsection{Reconstructed survival data}
\label{subsec:3:1}

KM curves are the richest source of information for AD meta-analysis. In the absence of IPD, it can be considered as a poor man's IPD if the survival data are properly reconstructed. The overview of a general strategy for data extraction is as follows.

The first step in reconstructing the survival data is to digitize the KM curves using sufficiently large, clear, and high-resolution images. In electronically published articles, images are typically either `raster images' or `vector objects'. A raster image consists of pixels, each represented by a dot or square with its own coordinates in a 2-dimensional grid. A graphical digitizing software, such as \textit{GraphClick} or \textit{DigitizeIt}, can be employed to convert and store the corresponding coordinates of the change points on the KM curves by clicking on the individual points using a mouse or automatic read-off \citep{Guyot2012}.  Alternatively, \cite{Liu2014} proposed directly reading in the lines of the PostScript file of a vector image, rather than using a digitizing software program to read in the coordinates from a raster image, and reported an improvement with 200 times better resolution.

After the curves have been digitized, the next step is to implement a numerical algorithm to elicit survival data. Early work in this area involved dividing the KM curves into several time intervals, which gives a good representation of event rates over time, while limiting the number of events (e.g., 20\%) within a given time interval \cite{Parmar1998}. It was later extended and implemented in a spreadsheet calculator by \cite{Tierney2007}. Using supplementary R code, \cite{Guyot2012} produced an iterative algorithm to derive a close approximation of the original individual patient time-to-event data from the published KM survival curves. Including the number of patients at risk and the number of events noted from the paper, if available, also significantly improved the approximation accuracy. \cite{Saluja2019} compared 4 data reconstruction algorithms \citep{Guyot2012, Hoyle2011, Parmar1998, Williamson2002}, and reported that the Guyot method was consistently superior to others.
	
Although the modeling of reconstructed survival data is lacking in the literature, the general ideas can be analogues to that of IPD without patient-level covariates. In IPD meta-analysis literature for survival outcomes, it was popular to apply a Poisson generalized linear model, which could be implemented with either fixed or random effects and with baseline hazard stratified by trial \citep{dear1994iterative, Arends2008}. \cite{Crowther2012} extended this model to allow non‐proportional hazards of the treatment effects. 

In MARS, we assume the reconstructed survival data follows a piecewise exponential distribution in the AD framework.  In theory, the piecewise exponential survival model can be equivalent to a Poisson log-linear model \citep{Holford1980, Laird1981}. To model the piecewise exponential, we first separate the overall time scale into $J$ prespecified disjointed small segments $I_j=(t_{j-1},t_j]$ for $j=1,2,\dots,J$, where $0=t_0<t_1<\dots<t_J<\infty$. For each patient, we create a set of pseudo-observations, one for each interval, indicating whether the individual had an event in that interval. The piecewise exponential model incorporates the data by treating the pseudo-observations as following Poisson distributions with means that are equal to the cumulative hazards in the intervals. Thus, the likelihood function of a piecewise exponential model is equivalent to the product of Poisson likelihood functions, one for each combination of an individual and an interval. 

Within each interval, the baseline hazard rate and the log HR for group difference are assumed to be constant, denoted by $\lambda_j$ and $\beta_j$ for $j=1,2,\dots,J$. We denote $t_J$ the largest survival or censored time. If some study has an extraordinary long follow-up period compared with other studies, the reconstructed survival time could be censored by a reasonable maximum time to make the inference of the meta-analysis meaningful and robust. 
The hazard rate for the $i$th subject in the $k$th study is modeled as
\begin{equation}\label{eq:1}
h_{k,i}(t)=h_{k}(t, X_{k,i})=\lambda_j\exp{\{\alpha_k+\theta_k(t)X_{k,i}\}},\quad t\in I_j,
\end{equation}
where $\alpha_k$ allows the baseline hazard rate to vary from study to study and $X_{k,i}$ indicates the treatment assignment. In the motivating case study (described in Section \ref{sec:2}), $X_{k,i}$ is the indicator of MRD status.
The study-specific log HR $\theta_k(t)$ is decomposed into the time-dependent effect $\beta_j$ and an inter-study effect $\mu_k$
\begin{equation}\label{eq:2}
\theta_k(t)=\beta_j+\mu_k,  \quad t\in I_j.
\end{equation}
The survival function is uniquely defined given the hazard function. If the study-level covariates (moderators) $Z_k$ are available, it is also straightforward to accommodate the study-level covariate effects, $\gamma Z_k$, into $\mu_k$ in (\ref{eq:2}) for subgroup analysis.

\subsection{Supplemental survival data}

One objective of MARS is to diminish selection bias. In meta-analysis applications, studies with survival information on intervention effect may not all publish KM curves, but report alternative summary statistics. To include all available information from eligible studies in a comprehensive systematic review, we incorporate into MARS with other common types of survival data, ranked by their richness of survival information. Without ambiguity, we denote the reconstructed survival data as the type I data hereinafter.

\subsubsection{Hazard ratios (type II data)}
\label{subsec:3:2} 


In MARS, HR estimates are synthesized as type II data if the KM curves are not available or are provided in low-resolution images, which is often the case in early publications. 
When the reported HRs are estimated under the proportional hazards assumption, they are presumed to be time-independent. Estimating a time-varying HR based solely on reported constant HRs is not feasible. However, given the various lengths of follow-up periods, it is natural to regard the observed constant log HR $\hat{\theta}_k$ as an estimate of the weighted average of stepwise time-dependent effects $\beta_j$ in addition to the study-specific effect $\mu_k$, 
\begin{equation*}
\hat{\theta}_k \sim N(\bar{\theta}_k, \hat{se}(\hat{\theta}_k)^2),
\end{equation*}
with the HR aligned with study duration
\begin{eqnarray*}
\bar{\theta}_k & = & \frac{1}{T_k} \int_{0}^{T_k}\theta_k(u)du\\
&=& \mu_k + \frac{1}{T_k}\{\sum_{l=1}^{j-1} \beta_l|I_l|+\beta_j(T_k-t_{j-1})\}, \quad T_k \in I_j,
\end{eqnarray*}
where $\theta_k(t)$ is defined in equation~(\ref{eq:2}), $|I_j|=t_j-t_{j-1}$, and $T_k$ is the duration of the $k$th study. The corresponding likelihood provides no information on the baseline hazard function.

In general, type II survival data can consist of the HRs converted from variants of indirect statistics comparing group differences in clinical studies lacking reported HRs. A typical example is a log-rank test statistics and its p-value, from which the log HR and its variance can be reasonably approximated \citep{Parmar1998}. 
This allows the totality of evidence to be assessed and improves the efficiency and reliability of meta-analysis for survival-type endpoints \citep{Tudur2001}.  
\cite{Tierney2007} reviewed the relevant methods that carefully manipulate published or other summary data to obtain HR summary statistics, classified by the type of information presented in trial reports. The resulting converted HRs can then be conveniently used as type II data in the MARS model.

\subsubsection{Survival probabilities (type III data)}
\label{subsec:4:3} 

Estimates of survival probabilities at particular time points (e.g., 1-year or 3-year DFS rates for individual groups) are often reported in randomized controlled clinical studies,  but usually ignored and rarely used in meta-analysis even if HR was not published.
Compared with HR estimates, survival rates provide limited information and thus are not considered valid surrogate measures for the meta-analysis of survival outcomes \citep{Michiels2005}. Although it can also be used to estimate HR estimates, such estimation requires assumptions in addition to the proportional hazards assumption. Nevertheless, it put an important piece to complete the information synthesis for systematic review when other types of survival data of a study are not available. For example, \cite{Berry2017} used it in a 2-step approach by first estimating the HR after assuming an exponential distribution of the time-to-event outcomes for the totality of evidence. 
	
In MARS, the reported survival rates are treated as the realizations of the survival function derived from equation (\ref{eq:1})
\begin{eqnarray}
\label{eq:3}
S_{k}(t,X)&=& \exp \left\{ -\int_{0}^{t}h_k(u,X)du\right\} \nonumber\\
&=&\exp \{-\sum_{l=1}^{j-1}\lambda_l\exp\{\alpha_k+(\mu_k+\beta_l) X\}|I_l| \nonumber\\
&& -\lambda_j\exp\{\alpha_k+(\mu_k+\beta_j) X\}(t-t_{j-1}) \}, \quad t\in I_j.
\end{eqnarray} 
When survival rate $\hat{S}_k(t,X=x)$ and its standard error $\hat{se}(\hat{S}_k(t,X=x))$ for $x$th treatment group are provided at a specific time point $t$ for the $k$th study, we assume that $\text{logit}(\hat{S}_k(t,X=x))$ follows a normal distribution as
\begin{equation*}
\text{logit}(\hat{S}_k(t,x))\sim N\left(\text{logit}(S_k(t,x)),\frac{\hat{se}^2(\hat{S}_k(t,x))}{\hat{S}_k(t,x)^2(1-\hat{S}_k(t,x))^2}\right)
\end{equation*}
to reduce the bias from asymmetry for interval-ranged $S_k(t,x)$ on $(0,1)$. $\hat{S}_k(t,x)=1$ or $0$ has been rarely reported, as the observation would be neglected to avoid the degenerate likelihood.

\subsection{Priors specification and model implementation}
\label{subsec:4:4} 

For the log HR estimation, $\beta_j$ represents the time-dependent treatment effect in time interval $I_j$ and $\mu_k$ describes the inter-study heterogeneity. We assume that prior of $\mu_k$ follows a normal distribution $\mu_k \sim  \mathcal N(\mu_0,\sigma_\mu^2)$. 
To specify the prior for $\boldsymbol\beta=(\beta_1,\beta_2,\dots,\beta_J)$, we rely on the multivariate normal prior: $$\boldsymbol\beta | \Sigma_{\boldsymbol\beta} \sim \mathcal{MVN}(\boldsymbol{\mu_\beta},{\Sigma_{\boldsymbol\beta}}),$$
where the mean $\boldsymbol{\mu_\beta}=\mu_\beta \mathbf{1}$, and the variance-covariance matrix $\Sigma_{\boldsymbol\beta}$ takes a first-order autoregressive structure
\begin{equation*}
\Sigma_{\boldsymbol\beta}=\sigma_\beta^2 \begin{pmatrix}
1 & \rho_\beta & \rho_\beta^2 & \dots & \rho_\beta^{J-1}\\
\rho_\beta & 1 & \rho_\beta & \dots & \rho_\beta^{J-2}\\
\dots & \dots & \dots & \dots & \dots \\
\rho_\beta^{J-1} & \rho_\beta^{J-2} & \rho_\beta^{J-3} & \dots & 1
\end{pmatrix}.
\end{equation*} 

The hyperparameter $\rho_\beta$ induces the association between the time segments and thereby the smoothness of the estimated piecewise HR function. For example, if $\rho_\beta=1$, this model implies that the time-dependent treatment effects are identical, that is, degenerate to a proportional hazards model.

The parameters $\lambda_j$ and $\alpha_k$ capture the baseline hazard rates and their variation among studies. Similarly, we assume a normal prior distribution for $\alpha_k$ as $\alpha_k\sim \mathcal N(\mu_\alpha,\sigma_\alpha^2)$, and a multivariate normal distribution for $(\log(\lambda_1),\log(\lambda_2),\dots,\log(\lambda_J))$ with the mean $\boldsymbol\mu_\lambda =\mu_\lambda \mathbf{1}$ and first-order autoregressive covariance matrix $\Sigma_{\boldsymbol\lambda}$ of the parameters $\sigma_\lambda^2$ and $\rho_\lambda$.  The hyperparameters follow the common choice of non-informative priors 
\begin{eqnarray*}
\mu_\beta,\mu_\lambda,\mu_\alpha,\mu_0 & \sim & \mathcal N(0,1000)\\
1/\sigma_\beta^2,1/\sigma_\mu^2,1/\sigma_\lambda^2 & \sim &  Gamma(0.01,0.01)\\
 \rho_\beta, \rho_\lambda & \sim &   Unif(-0.99,0.99).
\end{eqnarray*} 

For the case study, the joint posterior distributions were generated using Markov Chain Monte Carlo (MCMC) methods \citep{Gelman2003}. We used the R2jags package in statistical software R (version 3.6.1) to interface with the JAGS (Just Another Gibbs Sampler, version 4.3.0) for data analysis. More details can be found in the JAGS model specification in the Supplementary Materials.

\subsection{Summary measures for meta-analytic reporting}
\label{sec:3:5} 


We specify several summary measures, moving beyond the HR estimator, for statistical reporting and result dissemination using MARS. Having precise estimates of the difference in survival between treatments is extremely important, especially in oncology \citep{Michiels2005}. In conventional meta-analysis, HR estimates are routinely used to summarize differences in survival endpoints between groups \citep{Moher2010,Higgins2020}. 
If the underlying proportional hazards assumption is violated, the clinical interpretation of the HR estimate becomes difficult. Alternatively, owing to its flexible Bayesian model structure, MARS can generate more informative summary measurements to quantify the between-group difference with exact inference for uncertainty quantification.

\vspace{6pt}
\noindent\textbf{Survival probabilities.} \hspace{\parindent}
Survival probabilities are essential in survival analysis as they provide crucial summary information about the survival experiences of study populations. MARS readily yields the estimation of the overall survival function 
\begin{eqnarray}
\label{eq:4}
S(t) &=& S(t,X) \ = \ \exp \left\{ -\int_{0}^{t}h(u,X)du\right\} \nonumber\\
&=&\exp \{-\sum_{l=1}^{j-1}\lambda_l\exp\{\beta_l X\}|I_l| -\lambda_j\exp\{\beta_j X\}(t-t_{j-1}) \}, \quad t\in I_j.
\end{eqnarray} 

The Bayesian model can deliver exact inference on parameter estimation without assuming asymptotic normality.  Estimated survival probabilities can function similarly as KM estimates of the survival function so that treatment effects summarized in them are intuitive and straightforward. Using MCMC can generate the 95\% credible intervals (CrIs) for uncertainty quantification. Moreover, it is appealing to generate graphical representations of meta-analytic survival curves (see, for example, Figure 3A and 3B).

\vspace{6pt}
\noindent\textbf{Median survival time.} \hspace{\parindent}
The median survival time defines as the amount of time $t$ such that $S(t)=0.5,$ which is equivalent to that the cumulative hazard function $H(t)$ is $H(t)=\log(2).$

In MARS, the cumulative hazard is given by a piecewise linear function
\begin{equation*}
\hat{H}(t) = \sum_{l=1}^{j-1} \hat{\lambda}_l|I_l|+\hat{\lambda}_j(t-t_j),
\end{equation*}
where $t_{j-1}<t\leq t_j$. Subsequently, the median survival times $t_{MS}$ can be estimated solving of the following equation: $\hat{H}(t_{MS}) = \log(2).$

\vspace{6pt}
\noindent\textbf{Restricted mean survival time (RMST).} \hspace{\parindent}
The RMST is a well-established alternative measure of the “life expectancy” from the time of treatment to a specific time horizon of interest \citep{royston2013restricted}. The RMST is valid in survival models under any distribution, including the proportional hazards model. It is as powerful as the HR when the proportional hazards assumption is valid and even more powerful than the HR when the assumption is invalid. Another advantage of the RMST is that it primarily relies on the chosen time period so that it is less subject to follow-up time \citep{kim2017restricted}.

The RMST is defined as the expected value of survival time up to a pre-specified time point $\tau$. Let $T$ denote the time to the event and $S(t)$ denote the corresponding survival function. The RMST equals to the area under the survival curve from baseline to $\tau$. 
\begin{equation*}
RMST(\tau) = E(\min(T,\tau)) = \int_{0}^{\tau} S(u)du
\end{equation*} 
Because the survival curve as be approximated by KM estimates, the RMST can be nonparametrically estimated by using KM estimates of survival function \citep{Guo2019AnalyzingRM}.

In the proposed piecewise exponential model, the $RMST(\tau)$ can be approximated using the estimated mean survival probabilities $\hat{S}$:
\begin{equation*}
\widehat{RMST}(\tau) = \int_{0}^{\tau} \hat{S}(u,x)du =\sum_{i=1}^{j} \frac{\hat{S}(t_{i},X)-\hat{S}(t_{i-1},X)}{-\hat{\lambda}_{i}\exp{(X\beta)}},
\end{equation*}
where $\tau=t_{j}$ and $\hat{S}(t_i,X)=\exp(-\sum_{i=1}^j\hat{\lambda}_i\exp{(X\beta)}|I_i|)$. Using posterior samples from the MCMC algorithm, estimating the RMST and its standard deviation is succinct.

\section{Asymptotic properties}
\label{sec:5} 

Meta-analysis is widely used in the current big data era in all scientific areas for thorough statistical inference. As both the scale and number of individual studies increase, it is essential to understand the theoretical properties of MARS.

\subsection{Consistency}

To account for the heterogeneity among studies, MARS employs a hierarchical Bayesian structure in meta-analysis. Using a hierarchical Bayesian model provides consistent and accurate estimation \citep{rouder2005introduction}. Assume there are $K$ studies and $n_k$ subjects in the $k^{th}$ study. We denote $n$ as average sample size. Let ${\boldsymbol\beta}=(\beta_1,\beta_2,\dots,\beta_J)^T$ and ${\boldsymbol\eta}=(\eta_1,\eta_2,\dots,\eta_J)^T$ be the parameters of piecewise log HR and piecewise log baseline hazard rates, respectively, and denote $\hat{\boldsymbol\beta}$ and $\hat{\boldsymbol\eta}$ as the estimates generated by MARS. We denote $\mathcal{K}_1$, $\mathcal{K}_2$, $\mathcal{K}_3$, and $\mathcal{K}$ as sets of studies that provide reconstructed survival data (type I data), HRs (type II data), and survival rates (type III data), respectively. We have the following result:

\begin{theorem}
Under the regularity conditions (a)-(f) specified in Appendix A and for $k\in\mathcal{K}_1$, as the sample size $n_k \rightarrow \infty$ and the number of patients in each interval $n_{kj}$ becomes larger at a constant ratio $n_{kj}/n_k=c_j$, $j=1,\dots,J$;

Then, as the total number of studies $K \rightarrow \infty$ and the average sample size $n \rightarrow \infty$, $\begin{pmatrix} \hat{\boldsymbol\beta}\\\hat{\boldsymbol\eta} \end{pmatrix}$ converges to $\begin{pmatrix} \boldsymbol\beta\\\boldsymbol\eta \end{pmatrix}$ in probability.
\end{theorem}

According to Theorem 1, MARS yields consistent estimators for $\boldsymbol\beta$ and $\boldsymbol\eta$ with sufficient reconstructed survival data. As a result, the estimated summary measures, such as the overall survival rates, median survival time and RMSTs in \ref{sec:3:5},
are consistent as well.

\subsection{Relative efficiency}

The relative efficiency of conventional fixed-effect and random-effect AD meta-analyses has been rigorously studied and shown to be at most as efficient as IPD meta-analysis and asymptotically equivalent \citep{lin2010relative,chen2020relative}.  
Here, we extend the theoretical results and investigate the relative efficiency of MARS. Because survival rates (type III data) can only provide limited information, we are particularly interested in the efficiency of estimating the time-varying hazard ratio function by incorporating the commonly reported supplemental type II data into meta-analysis when reconstructed survival data are unavailable in some studies. 
Let $\hat{\boldsymbol\beta}^*$ and $\hat{\boldsymbol\beta}^*_{IPD}$ be the estimates generated from MARS and IPD meta-analysis using type I and II data, respectively. 

\begin{theorem}
Under the regularity conditions (a)-(f) specified in Appendix A, we have the following properties that
\begin{enumerate}
    \item for any fixed $K$ and $n$, 
    $Var(\hat{\boldsymbol\beta}^*)\geq Var(\hat{\boldsymbol\beta}^*_{IPD})$;
    \item if $Var(\hat{\boldsymbol\beta}_k)$ coincides with the variance of MLE of $\boldsymbol\beta_k$ for $k\in\mathcal{K}_2$, as $K,n\rightarrow\infty$ and $Kn^{-1/2}\rightarrow 0$, the asymptotic variances satisfy $$\lim_{\substack{n \to \infty, K \to \infty \\ Kn^{-1/2}\rightarrow 0}}
    Var(\hat{\boldsymbol\beta}^*)=\lim_{\substack{n \to \infty, K \to \infty \\ Kn^{-1/2}\rightarrow 0}} Var(\hat{\boldsymbol\beta}^*_{IPD}).$$
\end{enumerate}
\end{theorem}

According to the first inequality, $Var(\hat{\boldsymbol\beta}^*)- Var(\hat{\boldsymbol\beta}^*_{IPD})$ is positive semidefinite, indicating that the efficiency of MARS is at most the same as that of IPD meta-analysis. The second statement implies that asymptotically, MARS can achieve the full efficiency when the sample size increases at a higher order rate than the study number. Note that the condition of variance for type II data is met when the hazard ratio is constant. In particular, it holds even if hazard ratio is constant within the period of study duration $T_k$ and becomes time-varying. This might not be practical. Thus, we encounter acceptable efficiency loss in reality. 

\begin{corollary}
Let $\hat{\boldsymbol\beta}$ denote the estimates generated from MARS by using all 3 types of data, for fixed $K$ and $n$, $Var(\hat{\boldsymbol\beta})\leq Var(\hat{\boldsymbol\beta}^*).$
\end{corollary}

The corollary indicates that for a meta-analysis with finite number of studies, including studies that only reported type III data increases efficiency.

\section{\label{sec:5} Simulation}

We empirically evaluated the finite sample performance of the proposed meta-analysis method in 2 simulation studies.  First, we compared the MARS to conventional AD meta-analysis method and IPD meta-analysis in a setting in which the proportional hazards assumption is valid (\textbf{Case 1}). Second, we assessed the validity of MARS and its relative efficiency as compared to IPD meta-analysis when the proportional hazard assumption is invalid but two survival curves cross each other (\textbf{Case 2}). 
The performances of the methods are compared based on 1,000 replications.

\subsection{\label{sec:5:1} Simulation setting}
\noindent\textbf{Case 1.} \hspace{\parindent}
In each replication, we assumed 13 eligible studies and proportional hazards over time. The random effects for each study were generated as follows:
$$\alpha_k \sim \mathcal N(0,0.01), \mu_k \sim \mathcal N(0,0.1)$$
where $k=1,2,\dots,13$ and $X$'s are independent binary variables generated from a Bernoulli distribution with a rate of 0.55. Given $\alpha_k$ and $\mu_k$, the survival times in the $k^{th}$ study were independently generated from the hazard function 
$$h_k(t)=h_0(t)\exp(\alpha_k+(\mu_k+\beta)X)$$
where $h_0(t)$ is a Weibull distribution with shape 0.1 and scale 0.5 and $\beta=-0.6$ $(\exp(\beta)\approx0.55)$ is the time independent effect.  We assumed that survival times were right-censored according to an exponential distribution with a censoring rate of 0.01. The sample size for each study was randomly sampled from 50 to 150. The follow-up time for each study was randomly sampled from 80 to 120 months, at multiples of 5 months only. We equally partitioned the 120 months into 10 segments, each 12 months long.

To mimic the real-world scenario that HRs may not be reported in every clinical study, 
we assumed that 6 studies reported KM plots, 6 studies reported HR (including 3 studies with KM plots and 3 without), and 4 studies reported survival rates. For 3 studies that had both KM data and HR estimates, only the KM data were used for MARS, and only the HR estimates were used for AD random-effects model.  Therefore, we applied MARS to 6 studies of type I data, 3 studies of type II data, and 4 studies of type III data. In contrast, in the conventional AD meta-analysis, 6 studies were included to estimate the HR. Original survival data from all 13 studies were included in an IPD meta-analysis as a gold standard for comparison. 

We computed the log HRs using MARS, IPD meta-analysis and the conventional AD meta-analysis. 
For MARS and IPD meta-analysis, we also computed survival probabilities, median survival times and RMSTs at 1, 3, 5, and 10 years for each group. 

\noindent\textbf{Case 2.} \hspace{\parindent}
In each replication, we included 20 studies, for which $\alpha_k$, $\mu_k$ and $X$ were generated the same as in case 1. The survival times in each study were generated from the hazard function 
$$h_k(t)=0.01\times\exp(\alpha_k+(\mu_k+\beta(t))X)$$
where the log HR $\beta(t)=1-0.2^{(t/30-1)}$ is time-varying and for simplicity, the baseline hazard rate was constant over time. The censoring mechanism and the follow-up time were the same as those in case 1. 

We assumed that 9 studies reported KM plots, 12 studies reported HR (including 5 studies with KM plots), and 4 studies reported survival rates. Thus, in MARS, we accounted for 9 studies of type I data, 7 studies of type II data, and 4 studies of type III data. In contrast, in the conventional AD meta-analysis, 12 studies were included to estimate the HR. Data from all 20 studies were included in an IPD meta-analysis. 

As we did in case 1, we calculated survival rates, median survival times, and RMSTs at 1, 3, 5 and 10 years. Alternative to the point estimate of log HR, which is meaningless when the HR is not constant, we computed the average of the squared error between the log HR function $\beta(t)$ and the estimate log HR function $\hat{\beta}(t)$ over time, where $\hat{\beta}(t)$ would be a stepwise function in MARS.

\subsection{Results}

\begin{table}[ht]
\footnotesize
\caption{\label{tab:table1} Comparison of HR estimates. Abbreviations: CP, coverage probability of 95\% confidence interval (CI) or 95\% credible interval (CrI). }
\begin{center}
\begin{tabular}{llcccc}
    \hline
       &             & \multicolumn{4}{c}{log HR}                   \\ \cline{3-6} 
       &             & Mean (SD)   & MSE  & CP   & Length of CrIs \\ \hline
Case 1: proportional hazards \ & MARS & -0.63 (0.10) & 2.66 & 0.98 & 0.48           \\
       & IPD  & -0.64 (0.10) & 2.70 & 0.95 & 0.49           \\
       & AD   & -0.59 (0.12) & 2.55 & 0.96 & 0.46           \\
Case 2: time-varying hazards \ & MARS & 0.19 (0.10)  & 0.24 &      &                \\
       & IPD  & 0.23 (0.09)  & 0.19 &      &                \\
       & AD   & 0.05 (0.08)  & 1.48 &      &               \\ 
\hline
\end{tabular}
\end{center}
\end{table}

\begin{figure}[!ht]
\begin{center}
\includegraphics[width=\textwidth]{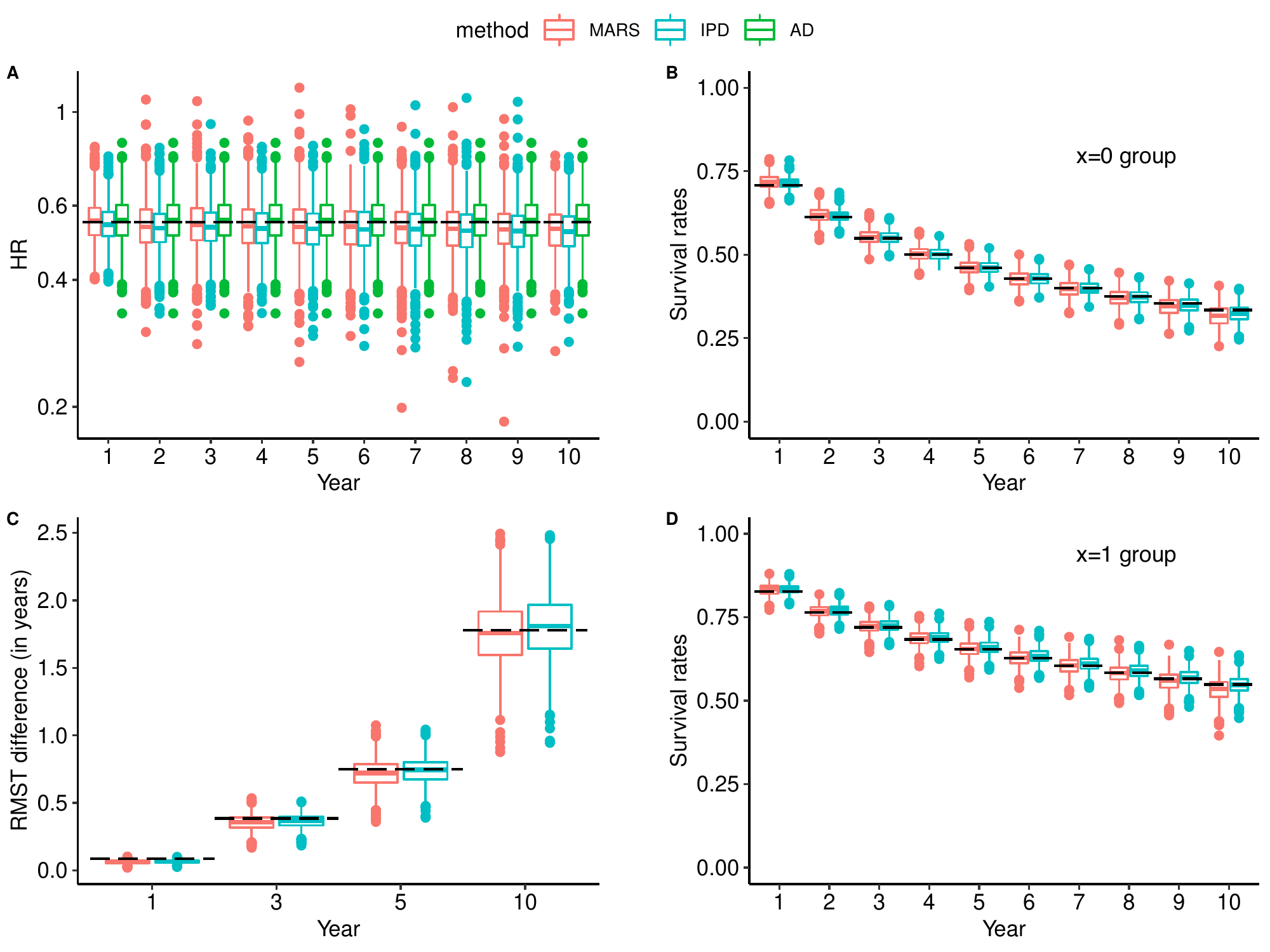}
\end{center}\vspace{-25pt}
\caption{\linespread{1.3}\selectfont{}\label{fig:FIG1}{Comparison of the performances of MARS and IPD meta-analysis in the simulation study under proportional hazard. (A) Estimated piecewise HRs; (C) Differences in RMSTs; (B) and (D) Survival probabilities of group $X=0$ and group $X=1$, respectively.}}\vspace{10pt}
\end{figure}

\noindent\textbf{Case 1.} \hspace{\parindent}
Figure \ref{fig:FIG1} shows the performance of MARS in which the proportional hazard assumption holds. The piecewise HRs estimated by MARS were approximately constant (Figure \ref{fig:FIG1}A), highlighting that MARS is able to identify a proportional hazards situation when the HR is actually constant. Thus, we compared the average piecewise HR estimates generated by MARS with the HR estimates generated by AD meta-analysis and IPD meta-analysis (Table \ref{tab:table1}).
Conventional AD meta-analysis using the available HR estimates from individual studies yielded the log HR estimates closest to the true value of -0.6. Although MARS generated a point estimate slightly smaller than the true value, the coverage probability of the 95\% CrIs was non-inferior to those of the other 2 methods. 

\vspace{10pt}

The estimated survival probabilities of both groupsgroup X=0 and group X=1  are shown in Figure \ref{fig:FIG1}B and \ref{fig:FIG1}D, respectively. MARS provided similar point estimates with slightly larger variability. Specifically, the true survival rates of the $X=0$ group at 3, 5, and 10 years were 0.55, 0.46 and 0.33, respectively (table \ref{tab:table2}). With MARS, the 95\% CrIs were more likely to cover the true survival rates at 3 years, 5 years and 10 years, whereas the IPD meta-analysis provided a more precise estimate of those survival rates.

\begin{table}[ht]
\caption{\label{tab:table2} Comparison of SR, MS and RMST. Abbreviations: CP, coverage probabitility of 95\% CrI; SR, survival rate; MS, median survival time.}\vspace{10pt}
\includegraphics[width=6.5in]{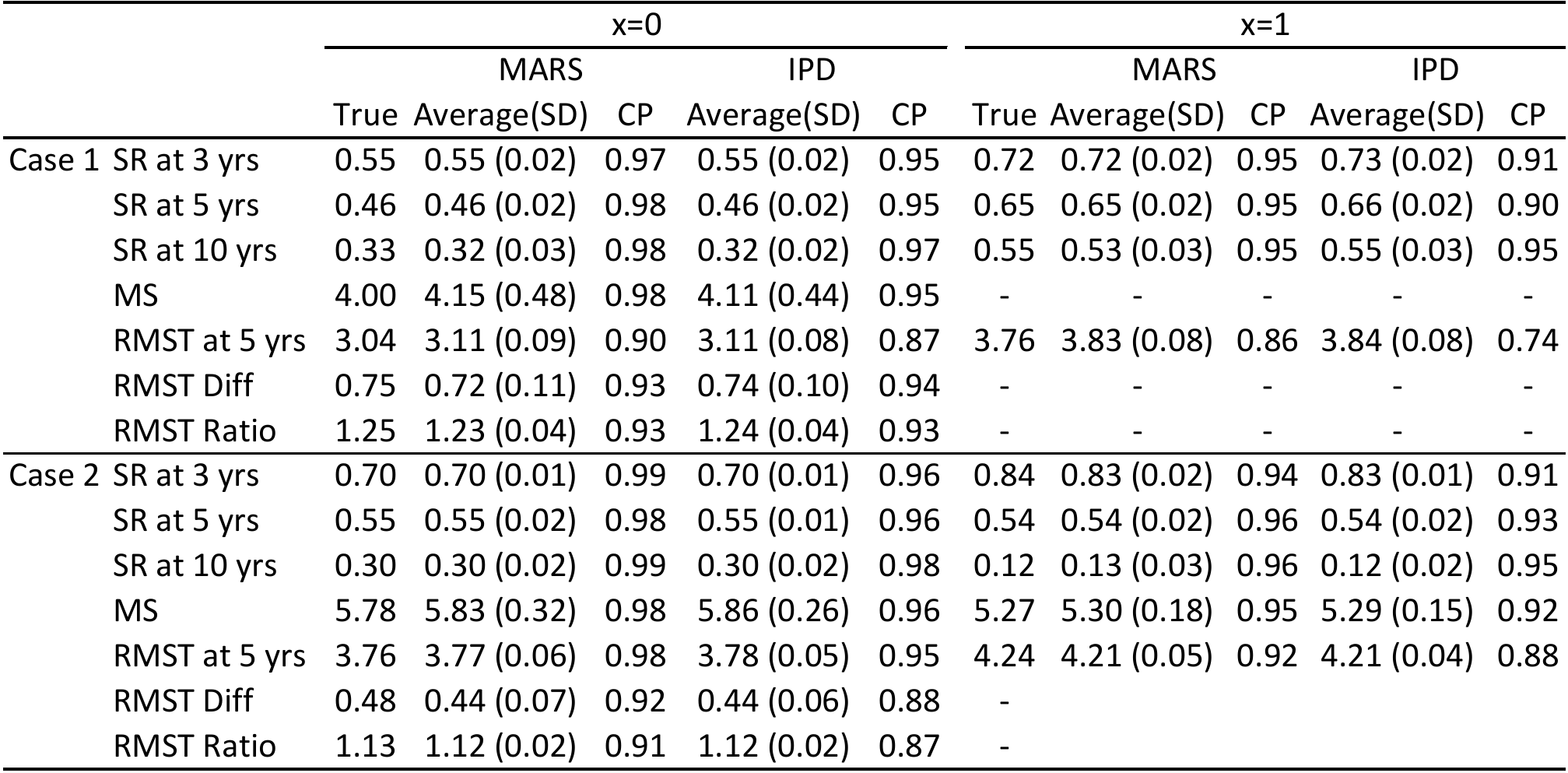}
\vspace{-45pt}
\end{table}

Regarding the median survival time and RMST (Figure \ref{fig:FIG1}C and Table \ref{tab:table2}), MARS had similar operating characteristics to those of the IPD analysis, and a higher coverage probability due to wider 95\% CrIs length. The mean relative efficiency of MARS to IPD meta-analysis was 1.22, indicating that MARS had reasonable efficiency loss compared with IPD meta-analysis.

\begin{figure}[ht]
\begin{center}
\includegraphics[width=\textwidth]{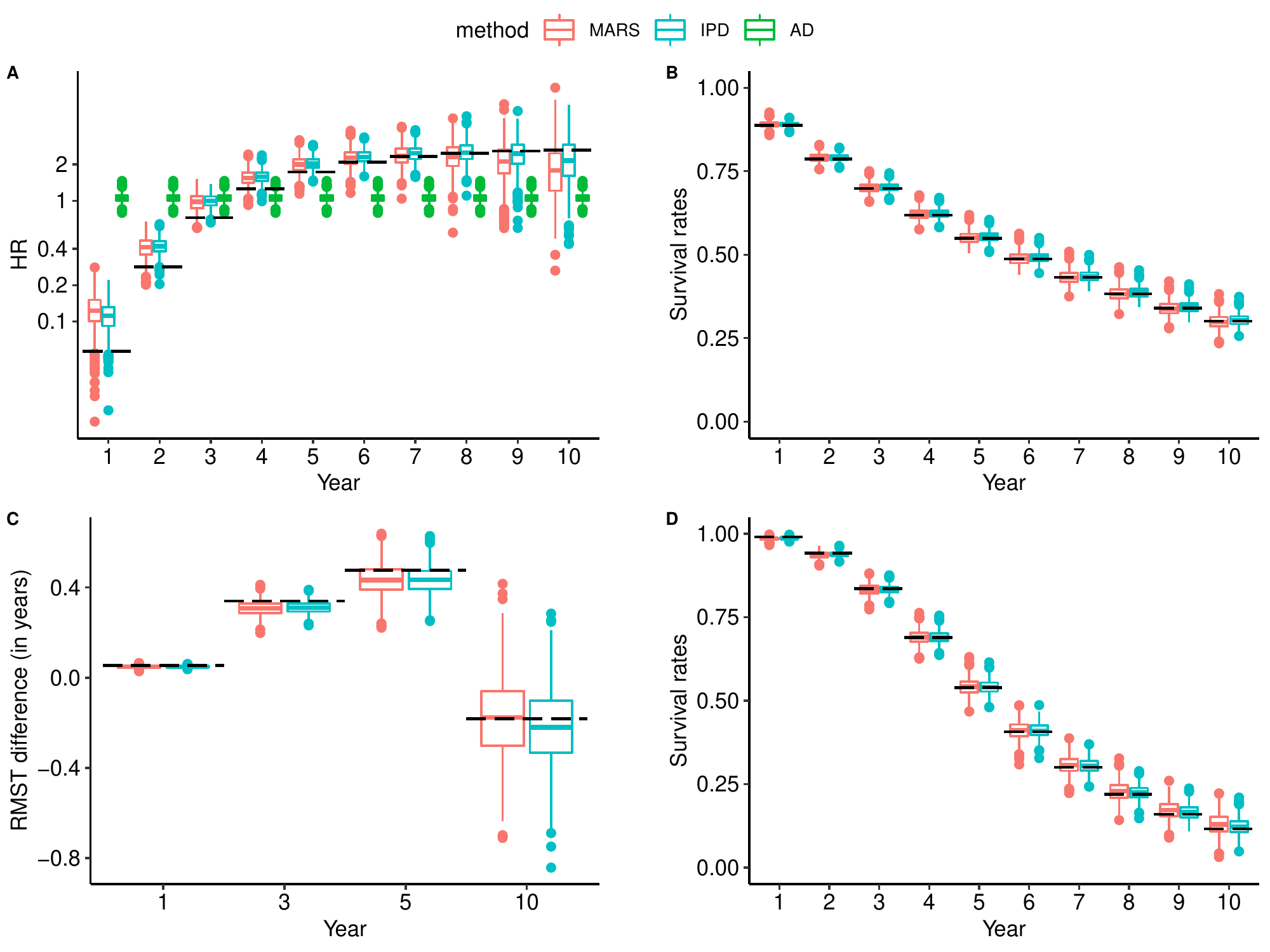}
\end{center}\vspace{-25pt}
\caption{\linespread{1.3}\selectfont{}\label{fig:FIG2}{
Comparison of the performances of MARS and IPD meta-analysis in the simulation study when non proportional hazard. (A)Estimated piecewise HRs. (C) Differences in RMST. (B and D) Survival probabilities of group $X=0$ and group $X=1$, respectively.}}\vspace{15pt}
\end{figure}

\noindent\textbf{Case 2.} \hspace{\parindent}
Figure \ref{fig:FIG2} demonstrates the performance of MARS using 3 types of data to provide estimates of various summary measures when the proportional hazards assumption is violated. The average of squared error of MARS (0.24) was higher than that of the IPD meta-analysis but much smaller than that of conventional AD meta-analysis, which is inappropriate in this situation (Table \ref{tab:table1}). 

MARS had good performance in estimating survival probabilities (Figure \ref{fig:FIG2}B and \ref{fig:FIG2}D). The survival rates MARS generated for both groups were almost identical to those generated from IPD meta-analysis. In long-term follow-up, both methods slightly over-estimated survival rates for group $X=1$. This can be explained by the shrinkage effect in Bayesian estimation of time-varying HR function \ref{fig:FIG2}A). Even though the HRs were over-estimated in early years, it effected less to the estimated survival rates due to the low hazard rates in that period. The 95\% CrIs generated with MARS were a little wider, but also had a higher probability to cover the true values at all time points (Table \ref{tab:table2}).

MARS was also able to give valid estimates of median survival time and RMSTs at 5 years with minor deviations from the true values (Table \ref{tab:table2}). 

\section{Application}
\label{sec:7} 

In our application of MARS to assess the MRD effects in AML dataset, we used time segments of 0-6 months and every 6 months of follow-up thereafter to a maximum of 11 years (22 intervals total).

\begin{figure}[ht]
\includegraphics[width=\textwidth]{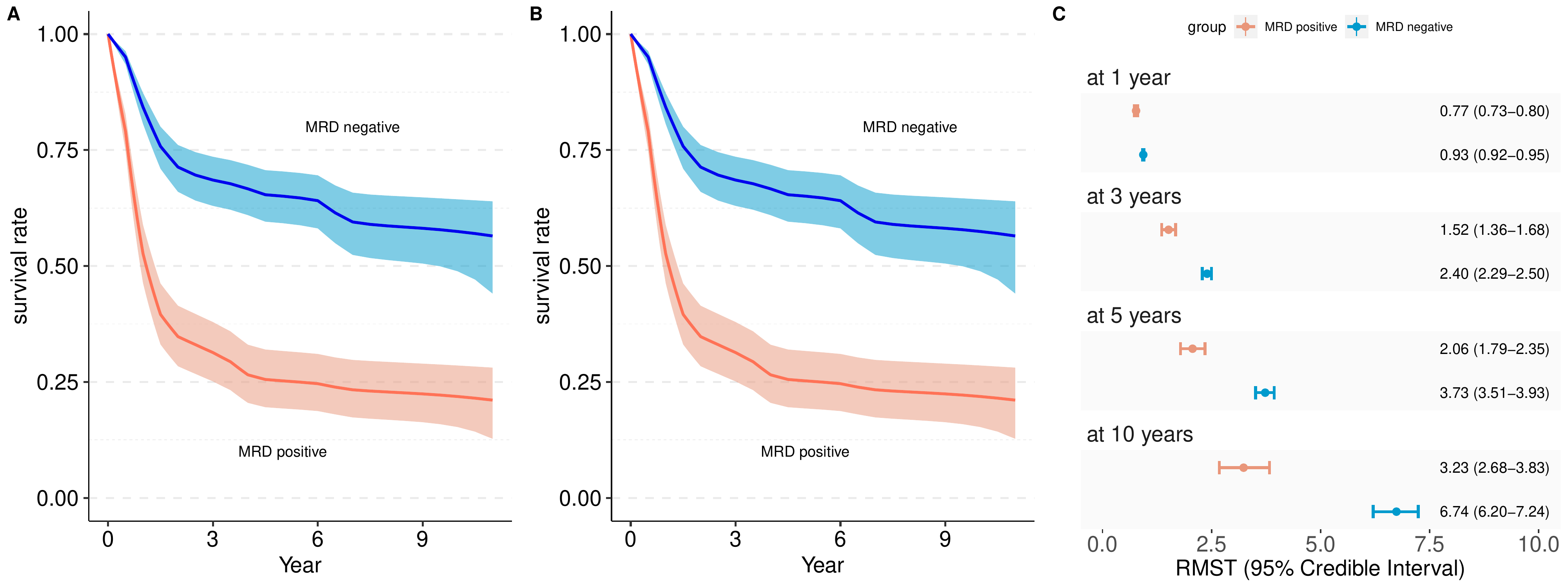}\vspace{-5pt}
\caption{\linespread{1.3}\selectfont{}\label{fig:FIG3}{Estimated DFS curves, stratified by MRD status, and estimated RMSTs. (A) Survival probabilities, with 95\% CrIs, estimated using all 3 types of survival data. (B) Survival probabilities, with 95\% CrIs, estimated using only reconstructed survival data. (C) Estimated RMSTs and corresponding 95\% CrIs for MRD-negative patients and MRD-positive patients at various times. 
}}\vspace{10pt}
\end{figure}

The meta-analysis survival curves for DFS are shown in the figure \ref{fig:FIG3}A, representing the survival probabilities over years. Generally, the DFS was better for subjects who showed MRD negative than those who had positive MRD status. The estimated DFS at 5 years were 65\% (95\% CrI, 59\%-70\%) for MRD negative group and 25\% (95\% CrI, 19\%-32\%) for MRD positive group. Figure \ref{fig:FIG3}B shows the survival curves for DFS synthesizing only reconstructed survival data (type I data) of 16 studies. Survival probabilities of both groups in figure \ref{fig:FIG3}B are lower than those in figure \ref{fig:FIG3}A, and the 95\% credible interval (CrI) was significantly narrower in figure \ref{fig:FIG3}B than CrI in figure \ref{fig:FIG3}A. Both discrepancies in point estimates and variability reflect the selection bias and extractor bias resulting from using only type 1 data and consequently suggest to conduct complete data analysis for heterogeneity estimation rather than using partial data, especially when the total number of studies included for meta-analysis is not large. The median survival time for MRD negative group is unavailable as the estimated survival probability at the end of study is about 0.56, while the median survival time for MRD positive group is around 1.08 years.

\vspace{20pt}


The RMSTs of both groups at 1, 3, 5, and 10 years are shown in Figure 3C. Patients who were MRD-negative had a longer RMST than patients who were MRD-positive.  The estimated RMST at 5 years was 3.73 years (95\% CrI, 3.51-3.93) for MRD negative group, while that for MRD positive group was 2.06 years (95\% CrI, 1.79-2.35).

\section{Discussion}
\label{sec:8}

In this work, we present MARS, a novel AD meta-analytic approach to exploit reconstructed survival data as a valuable source of information. 
Reconstructed survival data, which can be deemed as an AD alternative to IPD, enables theoretical advantages similar to using IPD in both meta-analytic modeling and reporting.
To systematically assess previous research studies, we develop a Bayesian multilevel modeling framework consolidating 3 regular types of survival information: 1) the reconstructed survival data elicited from published Kaplan-Meier curves, 2) log HR estimates, and 3) survival rates at specific time points. For the estimation of the HR function and survival probabilities, the piecewise exponential functions are employed with AR(1) correlations to model time-varying effects, the study effects are intrinsically weighted by the length of follow-up study duration, and MCMC allows exact inference for point estimation and uncertainty quantification.   
We establish the theoretical large sample properties, including consistency and relative efficiency, and provide conditions under which the MARS may coincide in efficiency with the IPD analysis. 
In finite sample, the MARS demonstrates comparable performance to IPD meta-analysis through simulation studies, under both the constant and time-varying hazard ratio scenarios.
As suggested by the motivating MRD example, this research endeavor can congruously
promote clinical research and assist regulatory decisions.

Meta-analysis modeling approaches for continuous and binary outcomes are relatively well developed, but less work has been done with time‐to‐event outcomes \citep{Freeman2017}.   Given the high prevalence of clinical studies in which the proportional hazards assumption is violated, the validity of AD meta-analysis is often in question.  The considerable popularity in meta-analysis of survival endpoints suggests the critical need for a modeling approach that can relax proportional hazards assumption \citep{Rulli2018}.  

The development of the MARS method contributes to the meta-analysis literature in several ways. 
In particular, MARS relaxes the proportional hazards assumption in individual clinical studies, which is a fundamental restriction in the meta-analysis using HR estimates \citep{Higgins2020}.
To derive valid inference for a given population, MARS comprehensively utilizes the most evidence available in the literature, and thereby mitigates the selection bias by including studies that have partial survival information.
Sensible summary measurements can be derived conveniently to enrich AD meta-analytic reporting for time-to-event endpoints. For example, survival probabilities can be generated for visual presentation to inform the risk of study group over time. Alternative statistics, such as survival probabilities and RMSTs, can be derived to better inform meta-analysis reporting especially if the proportional hazard assumption is not met. 
In summary, by addressing various limitations of current approaches, the MARS method effectively extends the classical framework of AD meta-analysis for survival endpoints.

Only 25\% of published IPD meta-analyses have had access to all IPD \citep{Nevitt2017}.  A generalization of the MARS method can be applied to combine IPD and AD data for survival endpoints \citep{riley2007evidence}, or process information together with single-arm trials or multiple treatments to incorporate real-world evidence \citep{begg1991model, zhang2019bayesian}. Additionally, one future methodological exploration is to employ alternative strategies for estimating time-varying effects. \cite{Jansen2011, Jansen2012} used fractional polynomials \citep{Royston1994} to model the time-varying hazard functions in a two‐step IPD analysis of time‐to‐event data.  Fractional polynomials are a parsimonious alternative to regular polynomials and provide flexible parameterization for continuous variables.  However, the shape of a fractional polynomial may result in unexpected end effects which may jeopardize the overall accuracy. \cite{Freeman2017} applied the Royston‐Parmar model with the baseline log‐cumulative hazard modeled by restricted cubic splines.  However, their formulation failed to guarantee the monotonicity of the cumulative hazard function.  Possibly, one solution is to adopt the extended multiresolution hazard model \citep{Hagar2017} for reconstructed survival data.   




\vspace{35pt}
\begin{center}
{\large\bf SUPPLEMENTARY MATERIAL}
\end{center}

\begin{description}

\item[Title:] Supplementary Materials for "Evidence synthesis with reconstructed survival data". (PDF file)

\end{description}

\newpage
\bibliographystyle{agsm}
\bibliography{MARS.bbl}

\end{document}